\documentclass[10pt,twocolumn,letterpaper]{article}
\usepackage{soul}
\usepackage[accsupp]{axessibility} 
\usepackage{booktabs}
\usepackage{multirow}
\usepackage{array}
\usepackage{graphicx}
\usepackage[table, dvipsnames]{xcolor}
\usepackage{microtype}

\definecolor{Trd}{rgb}{0.95,0.95,0.65}
\definecolor{Snd}{rgb}{1, 0.85, 0.7}
\definecolor{Fst}{rgb}{1, 0.7, 0.7}

\usepackage{iccv}      

\newcommand{\boldparagraph}[1]{\vspace{0.1cm}\noindent{\bf #1}}

\makeatletter
\DeclareRobustCommand\onedot{\futurelet\@let@token\@onedot}
\def\@onedot{\ifx\@let@token.\else.\null\fi\xspace}

\def\etc{etc\onedot}

\def\etal{et~al\onedot}

\makeatother

\newcommand{\figref}[1]{Figure~\ref{#1}}

\newcommand{\eqnref}[1]{Eq.~\ref{#1}}
\newcommand{\tabref}[1]{Table~\ref{#1}}

\definecolor{iccvblue}{rgb}{0.21,0.49,0.74}

\usepackage[pagebackref,breaklinks,colorlinks,citecolor=iccvblue]{hyperref}

\def\paperID{8942} 
\def\confName{ICCV}
\def\confYear{2025}

\title{Neural Shell Texture Splatting: More Details and Fewer Primitives}

\author{
Xin Zhang\textsuperscript{1} \quad
Anpei Chen\textsuperscript{2} \quad
Jincheng Xiong\textsuperscript{1} \quad
Pinxuan Dai\textsuperscript{1} \quad
Yujun Shen\textsuperscript{3} \quad
Weiwei Xu\textsuperscript{1}\thanks{Corresponding author}\\
\textsuperscript{1}Zhejiang University \quad
\textsuperscript{2}Westlake University \quad 
\textsuperscript{3}Ant Group \\
\url{https://zhangxin-cg.github.io/nest-splatting/}
}

\begin{document}
\maketitle
\begin{abstract}
Gaussian splatting techniques have shown promising results in novel view synthesis, achieving high fidelity and efficiency. However, their high reconstruction quality comes at the cost of requiring a large number of primitives. We identify this issue as stemming from the entanglement of geometry and appearance in Gaussian Splatting.  
To address this, we introduce a neural shell texture, a global representation that encodes texture information around the surface. We use Gaussian primitives as both a geometric representation and texture field samplers, efficiently splatting texture features into image space.  
Our evaluation demonstrates that this disentanglement enables high parameter efficiency, fine texture detail reconstruction, and easy textured mesh extraction, all while using significantly fewer primitives.
%
\end{abstract}

\section{Introduction}
\label{sec:intro}
Novel View Synthesis (NVS) generates images from new camera angles that are plausibly consistent with a set of conditioning images, allowing broad applications in virtual reality, robotics, and digital content creation. Among the most influential advancements in this domain is Neural Radiance Fields (NeRF)~\cite{mildenhall2020nerf}, which represents 3D scenes using a neural network for photorealistic novel view generation. Despite its success, NeRF suffers from computational inefficiency, motivating the search for faster alternatives. Recently, 3D Gaussian Splatting (3DGS)~\cite{kerbl3Dgaussians} has emerged as a compelling solution, offering real-time, high-quality rendering by leveraging efficient Gaussian primitives. This breakthrough has sparked significant interest in extending 3DGS to dynamic scenes, geometry, anti-aliasing techniques, and generative 3D modeling, pushing the boundaries of NVS toward practical deployment.

Specifically, 3DGS represents complex scenes using a set of 3D Gaussian primitives, which are efficiently rendered onto the screen via splatting-based rasterization. Each Gaussian is defined by attributes such as position, size, orientation, opacity, and color, all of which are stored independently and optimized using a multi-view photometric loss to accurately reconstruct scene appearance. 
A promising follow-up, 2DGS~\cite{Huang2DGS2024}, replaces the 3D Gaussian representation with 2D oriented planar Gaussian surfels and introduces a backward ray tracing technique for ray-splat intersection and rasterization, leading to more precise geometry modeling. 

However, the high rendering quality and accurate surface reconstruction of these methods come at the cost of requiring a large number of Gaussian primitives. This is due to the inherent coupling of geometry and appearance in their primitive representation, which requires significant densification in regions with complex textures or detailed geometry. 
In practice, the frequency of surface textures is often much higher than that of geometry. Therefore, prior GS-like methods suffer from underparameterization for texture and overparameterization for geometry.

\begin{figure}[t] 
\centering
\includegraphics[width=1\linewidth]{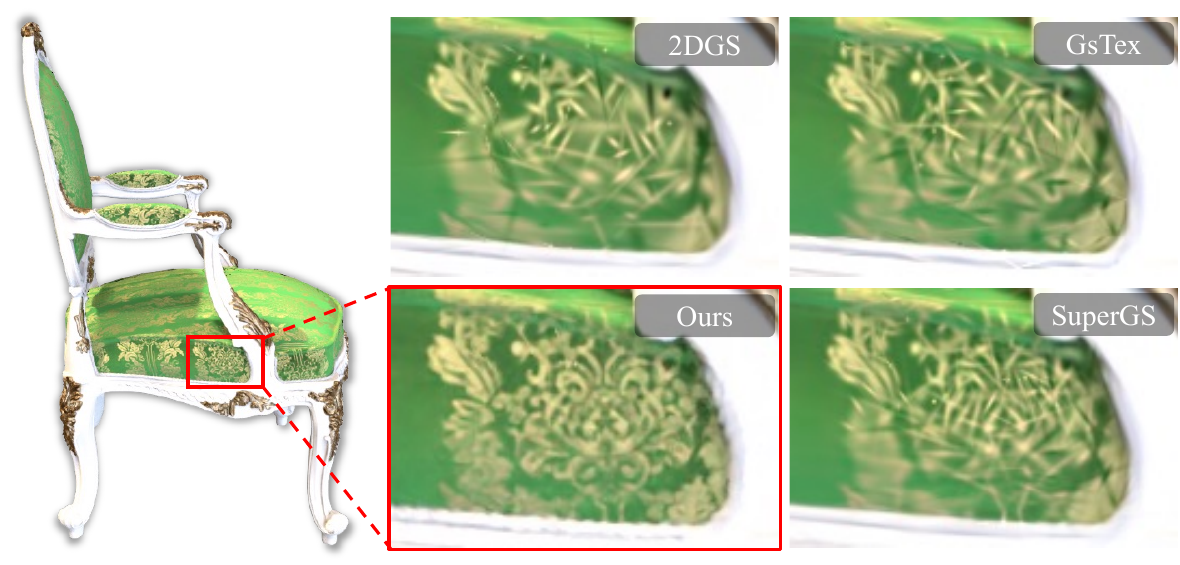}
\caption{Our method achieves enhanced details in rendering by replacing the splat-constant, view-dependent color in Gaussian Splatting with a global shell texture.}
\label{fig:teaser} 
\end{figure}

To resolve these issues, we propose to disentangle geometry and appearance in a novel way: the high-frequency texture is represented and shared across different positions using a global neural shell texture, while geometry can be modeled with 2D surfels for efficiency. The texture and geometry are then separately represented, and they are connected and rendered to images using the ray-splat intersections. This is different from previous works~\cite{xu2024supergaussians, rong2024gstex, chao2024texturedgs, svitov2025billboardsplatting} that employ spatially varying textures for each Gaussian primitive. These methods still associate textures to 3D points and  fail to significantly reduce the number of primitives while maintaining high-quality reconstructions, since the higher texture capability provides a highly discontinuous gradient field, leading to unreliable optimization.
We adopt Instant-NGP~\cite{muller2022instant} to model neural shell texture in a canonical representation space, while using Gaussian primitives as explicit geometry samplers.
%
In addition, we model and query geometry using efficient splatting instead of sampling points along the rays like in NeRF.
Consequently, our method is able to reconstruct objects with high-frequency texture details using $3$x fewer primitives and improves the rendering results at high-frequency texture areas, as shown in Fig.~\ref{fig:teaser}. 

In summary, our contributions include:

\begin{itemize}[leftmargin=12pt,topsep=4pt]
\item We propose \textbf{Ne}ural \textbf{S}hell \textbf{T}exture Splatting (NeST-Splatting), enabling Gaussian primitives with spatially varying texture colors, which significantly enhances the ability to represent fine details.

\item We disentangle texture and geometry in Gaussian Splatting, representing scenes with fewer primitives and a more compact implicit appearance model.

\item Our method naturally mitigates the needle-like artifacts commonly observed in Gaussian-based methods by enabling low-frequency geometry modeling through texture-geometry decoupling.

\end{itemize}

Experiments demonstrate that our method achieves state-of-the-art performance on the NVS task, exhibiting superior rendering quality in texture-rich regions.
\section{Related Work}
\label{sec:formatting}

\boldparagraph{Novel View Synthesis.}
Given multi-view images, novel view synthesis generates images for unseen viewpoints.
Previous methods achieve this by reconstructing light fields~\cite{Davis_Unstructured, Levoy_Light} or blending nearby views  based on geometry proxies~\cite{Gortler_lumigraph, Buehler_Unstructured, Hedman_Scalable, DeepBlending2018, Chen2018I3D}.

Neural Radiance Fields (NeRF)~\cite{mildenhall2020nerf} revolutionized the field of novel view synthesis by optimizing a neural scene representation using only photometric supervision. 3D coordinates with frequency encoding are fed into multi-layer perceptrons (MLPs) to predict geometric opacity and appearance color, aggregating in differentiable volumetric rendering.
Following works improve NeRF on anti-aliasing rendering~\cite{Barron_2021_ICCV, Barron_2022_CVPR}, reflection modeling~\cite{verbin2022refnerf}, large-scale reconstruction~\cite{Turki_2022_CVPR, Tancik_2022_CVPR, wu2023scanerf}, acceleration with grid features~\cite{yu_and_fridovichkeil2021plenoxels, Chen2022ECCV, mueller2022instant, SunSC22} \etc. 
Instant-NGP~\cite{mueller2022instant} employs multi-resolution hash grid and small MLPs to significantly speedup both the training and rendering of NeRF.
The hash grid flats 3D grid features into a 1D array using the corresponding 3D coordinates with a hash projection mapping.
It eliminates the redundancy of unused features in space and enables flexible compression rate control by the hash table length.
Compact-NGP~\cite{takikawa2023compact} reduces storage requirements by using smaller hash tables indexed by learnable probes, at the cost of longer training times. However, real-time rendering remains challenging for NeRF-based methods due to the hundreds of MLP queries needed per pixel.

3D Gaussian Splatting (3DGS)~\cite{kerbl3Dgaussians} represents scenes explicitly using anisotropic 3D Gaussian kernels with spherical harmonic (SH) color coefficients. By leveraging tile-based rasterization, it efficiently renders and optimizes Gaussian kernels through pixel-wise primitive blending. 3DGS achieves state-of-the-art novel view synthesis quality while maintaining over 100 FPS, even on low-end GPUs.  
Mip-Splatting~\cite{Yu2023MipSplatting} addresses aliasing issues in 3DGS by applying 3D smoothing and 2D mip filters. 2DGS~\cite{Huang2DGS2024, Dai2024GaussianSurfels} flattens ellipsoidal 3D Gaussian kernels into elliptical disks for more accurate geometry modeling. Other advancements improve rendering quality by using more expressive kernel functions~\cite{yu20242dgh2dgh, huang2024deformable} or enhanced densification strategies~\cite{scaffoldgs, Mallick_Taming, zhang2024pixelgs}.  
However, the explicit nature of 3DGS requires storing a large number of per-Gaussian parameters, leading to high storage overhead as the number of points increases. Efforts to reduce model size include quantization techniques~\cite{navaneet2023compact3d, lee2024c3dgs} and replacing SH with hash grid assisted appearance models~\cite{hac2024, lee2024c3dgs}. Another line of work~\cite{ye2024gsdr,DeferredGS,zhang2024ref} leverages deferred rendering to handle high reflective surfaces, achieving promising results.

Our method leverages the best from both worlds, using Gaussian primitives for geometry modeling and a continuous neural field for texture. 
We query the texture field at exact ray-Gaussian intersection points. 
This fully disentangles shape from texture, enabling higher-quality appearance with fewer Gaussian primitives.


\begin{figure*}[!t] 
\centering
\includegraphics[width=0.95\linewidth]{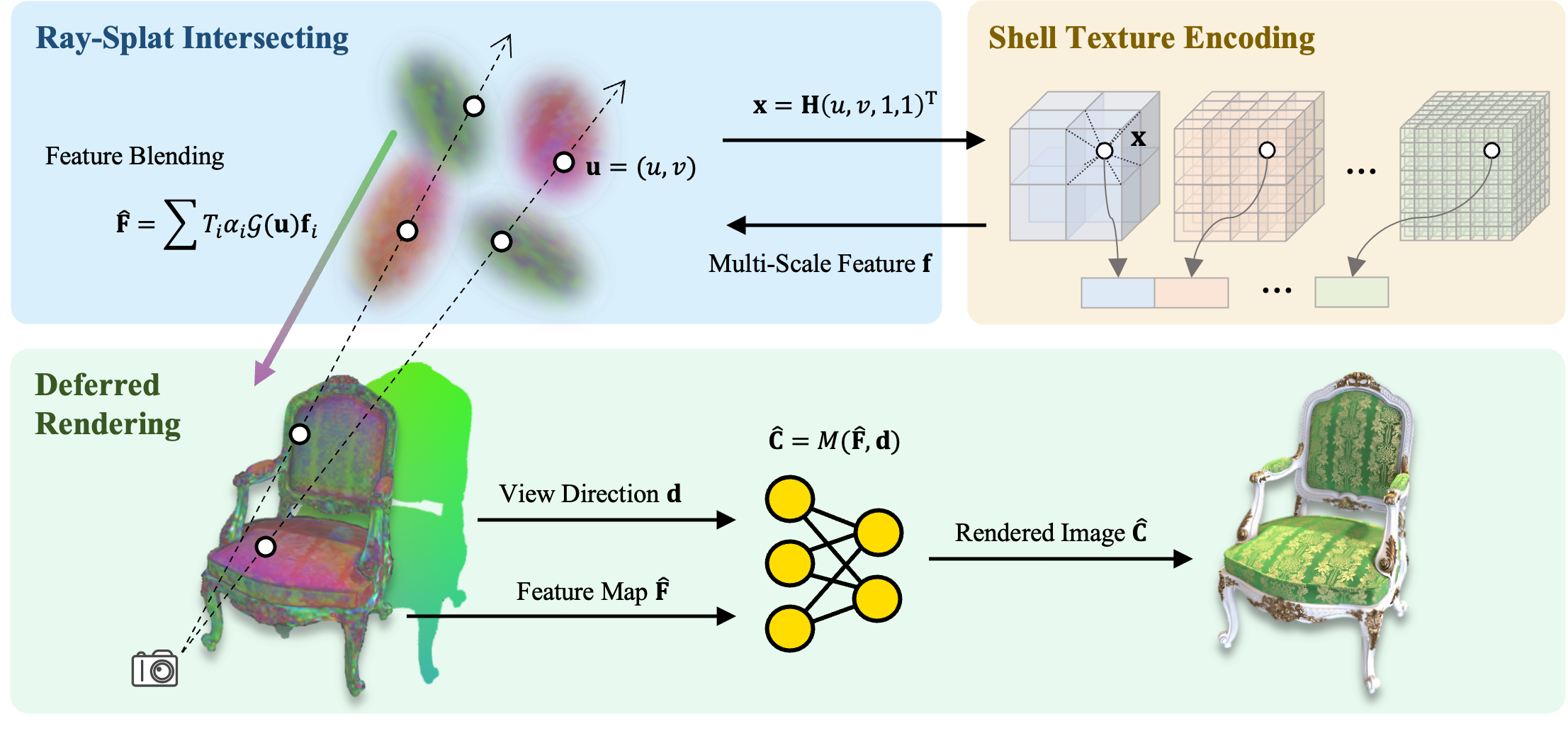}
\caption{\textbf{Overview of our method.} 
We query the hash grid encoded shell texture by ray-splat intersections.
The multi-resolution features are alpha-blended as screen-space feature map to perform deferred rendering efficiently. Our method fully disentangles the scene geometry and appearance, improving rendering quality on complex textures using fewer Gaussian primitives.
}
\label{fig:pipeline} 
\end{figure*}

\boldparagraph{Texture Representation and Reconstruction.}
Classic texture mapping~\cite{Levy_Least, Botsch_Geometric} applies 2D textures to mesh surfaces based on corresponding UV atlas.
A desirable property of texture mapping is the disentanglement of appearance and geometry, allowing the texture resolution to remain independent from the geometry complexity.
Recent approaches~\cite{henzler2020neuraltexture, Thies_Deferred, xiang2021neutex} incorporate neural network to parameterize 2D texture implicitly.
Volumetric textures~\cite{Kajiya89Rendering, Neyret98Modeling} are designed to model mesostructures like leaf and fur, which can also be efficiently parametrized by NeRF~\cite{huang2023nerf-texture, baatz2022nerf,Esposito2025CVPR,adaptiveshells2023}.

3DGS~\cite{kerbl3Dgaussians} assigns splat-constant parameters to each Gaussian point, tightly coupling SH-encoded appearance with geometry. This entanglement leads to storage overhead in two key aspects:
1) Appearance redundancy. Similar SH copies are stored independently in local regions where geometry varies more than appearance.
2) Geometry redundancy. Excessive Gaussian points are cloned in texture-rich regions where the geometry is rather simple.

Recent works empower Gaussian points with spatially varying colors to ease the geometry redundancy using learnable UV mapping~\cite{xu2024texturegs}, per-primitive texture map~\cite{rong2024gstex, svitov2025billboardsplatting, chao2024texturedgs}, per-primitive tiny MLPs~\cite{xu2024supergaussians}, or SH based on ray intersections~\cite{huang2024texturedgs}.

Although these methods enhance texture representation, they often learn homogeneous colors with slight variations due to optimization challenges in practice. Additionally, each Gaussian point is stored independently thus incurring storage overhead.
Instead of modeling texture locally, our method represents it as a neural shell around the primitives, queried via ray-splat intersections. This results in a continuous, semantically meaningful texture field and allows for easy textured mesh extraction.









\section{Method}


\subsection{Preliminaries}

\paragraph{Gaussian Splatting}
3DGS~\cite{kerbl3Dgaussians} employs Gaussian primitives to represent scenes.  
Each Gaussian point is associated with a group of parameters $\{ \mathbf{p}_i,\mathbf{S}_i, \mathbf{R}_i,\alpha_i,$ $\mathbf{SH}_i \}$, denoting the center position, 3D scaling, rotation, opacity and SH-encoded color.
The Gaussian parameters are optimized through differentiable volume splatting using highly parallelized rasterization.
Specifically, a 3D covariance matrix \!$\mathbf{\Sigma}$ is formed from $\mathbf{S}_i$ and $\mathbf{R}_i$, which defines the shape and orientation of an ellipsoid centered at $\mathbf{p}_k$.
For efficient rendering, the \!$\mathbf{\Sigma}$ is projected onto screen space, resulting in a 2D covariance matrix described by $\mathcal{G}^{2D}$.

To render the color of pixel $\mathbf{x}$, 3DGS employs point-based $\alpha$-blending to integrate the appearance of each Gaussian from front to back:

\vspace{-2.0em}
\begin{equation}
\begin{split}
    \mathbf{c}(\mathbf{x})\!=\! \sum_{i=1}^{n} T_i \alpha_i \mathcal{G}^{2D}_i(\mathbf{x})\mathbf{c}_i,  \hspace{0.1cm}  T_i\!=\!\prod_{j=1}^{i-1} (1\!-\!\alpha_j \mathcal{G}^{2D}_j(\mathbf{x})),
\end{split}
\end{equation}
where $i$ is the index of Gaussians intersected by the ray $\mathbf{x}$ and $ \mathbf{c}$ is the SH color of Gaussian Primitives. 

Building on top of 3DGS, 2DGS~\cite{Huang2DGS2024} replaces the 3D Gaussian representation with 2D oriented planar Gaussian disks (surfels) by reducing 3D $\mathbf{S}_i$ to 2D.
It also introduces a backward ray tracing technique for ray-splat intersection and rasterization, resulting in a more efficient way to model geometry and compute precise projections. 

The local tangent plane of each primitive in world space is therefore defined as:
\begin{equation}
    P(u,v)= \begin{bmatrix}\mathbf{RS} & \mathbf{p}_i \\0 & 1\end{bmatrix}(u,v,1,1)^{\mathrm{T}} = \mathbf{H}(u,v,1,1)^{\mathrm{T}},
\end{equation}
where $\mathbf{H}\!\in\!4\!\times\!4$ is a homogeneous transformation. The Gaussian intensity of point $\mathbf{u} = (u,v)$ in local tangent plane is defined as: $\mathcal{G}(\mathbf{u}) = \mathrm{exp}(-\frac{u^2+v^2}{2})$.
The local $uv$ coordinates are then projected to screen space for volume rendering, following the same procedure as in 3DGS.


\paragraph{Multi-resolution Hash Encoding.}
To address the high computational costs of NeRF, M\"uller \etal introduces Instant-NGP ~\cite{muller2022instant}, which augments a small neural network with a multi-resolution hash table of trainable feature vectors. More specificlly, Instant-NGP defines multi-resolution voxel grids in a bounded space, where each voxel grid is mapped to a set of 1D feature vectors via a hash function $h: \mathbb{Z}^d \to \mathbb{Z}_{\mathrm{T}}$. For a 3D position $\mathbf{x} \in\mathbb{R}^{3}$, hash encoding interpolates feature vectors from voxel grids containing $\mathbf{x}$ at each resolution level, followed by concatenating features across all levels:
\begin{equation}
    \mathbf{f}(\mathbf{x})=\{\mathbf{f}^{i}(\mathbf{x})\}_{i=0}^{L} \in \mathbb{R}^{L\times F},
    \label{eq:feature_query}
\end{equation}
where $\mathbf{f}^{i}(\mathbf{x})$ is the hash encoding at level $i$, $L$ is the number of levels, $T$ is the hash table size, $F$ is the feature dimension per level.
With trainable multi-resolution features and an MLP adaptively mitigating the influence of hash collisions, the multi-resolution hash encoding method captures the most important details, achieving high rendering quality while maintaining a compact structure.


\subsection{Modeling}

Our method decouples the geometry and appearance of Gaussian Splatting, allowing fewer primitives to represent scenes with complex appearance.
We propose to replace the SH modeling of Gaussian primitives with a global shell texture, parameterized by multi-resolution hash grid. 
As shown in \cref{fig:pipeline}, our shell texture Splatting employs 2D Gaussian Primitives for geometry representation and multi-level neural features for appearance modeling. 
This shell texture predicts a spatially varying color for each sample point, thereby achieving superior rendering in texture-rich regions without excessive point splitting or producing \emph{needle-like} artifacts.


\noindent \textbf{Appearance Modeling.} 
In Gaussian Splatting, the color of each splat is represented by a spherical harmonics function ($S\!H$):
\begin{equation}
    \mathbf{c}_i = S\!H(\mathbf{d}, \mathbf{SH}_i),\hspace{0.2cm}\textrm{where} \,\, \mathbf{d}=\mathbf{p_i} - \mathbf{o} ,
\end{equation}
where $\mathbf{o}$ is the camera position and $\mathbf{d}$ is the view direction, thus providing a splat-constant color related to the viewing direction. Note that the geometry and appearance are entangled, as the color field around the primitive is influenced by its shape in terms of position, orientation, and scale.
This entanglement requires a large number of tiny Gaussian points to capture color variations in complex textures. Moreover, since Gaussian kernels decay their weights smoothly, they shrink into \emph{needle-like} shapes to represent sharp texture edges, which degrades rendering quality in novel views.

To address this issue, we integrate Gaussian Splatting with a learnable, locally continuous, thin texture field over the entire 3D space, termed \emph{shell texture}, for spatially varying appearance on 2D splat surfaces.
Specifically, for each ray-splat intersection $(u,v)$, we use multi-resolution features $\mathbf{f}(\mathbf{p}_{\mathbf{u}})$ to represent its color, obtained by querying the corresponding world space point $\mathbf{p}_{\mathbf{u}}$ from a hash grid \eqnref{eq:feature_query}.

Despite its simplicity, our method provides a fundamental adjustment compared to prior work. Namely, it achieves full decoupling of geometry and appearance, representing the scene through a combination of explicit Gaussian point geometry and a compact neural texture field, instead of binding per-Gaussian color to each primitive. 
With our modeling, the Gaussian shape is independent of texture complexity and only serves as planar samplers for the implicit appearance field, allowing for flexible control on capacity of texture and geometry independently. 


In addition, the decoupling also enables a more compact appearance representation, without the post-quantization that is commonly needed in other compression GS methods~\cite{navaneet2023compact3d, lee2024c3dgs}. Since hash encoding takes the world coordinates of intersection points as input, nearby Gaussian primitives and those that map to the same 1D position via the hash function can share the same hash features efficiently. 
Note that our decoupled design supports various coordinate-conditioned appearance models. We choose Instant-NGP here for its simplicity and efficiency.


However, in Instant-NGP, features are decoded into RGB colors before volume rendering. With $K$ points sampled along each pixel ray, the decoding batch size becomes $K\!\times\!H\!\times\!W$, resulting in significant computational overhead and slow rendering. To avoid the computational cost of decoding features at every ray-splat intersection in our method, we employ deferred neural rendering by:
%
\begin{equation}
\begin{split}
\hat{\mathbf{F}} = \sum_{}^{} T_i \alpha_i \mathcal{G}(\mathbf{u_i})\mathbf{f}_i, &\hspace{0.5cm} 
\mathbf{C} = M(\hat{\mathbf{F}}, \mathbf{d}) ,
\end{split}
\end{equation}
where $\hat{\mathbf{F}}\!\in\! \mathbb{R}^{(LF) \times H \times W}$ denotes the feature image, $M$ refers to a small MLP decoder.
By integrating alpha-weighted features from front to back, we obtain a feature image $\hat{\mathbf{F}}$, which is then decoded by $M$ into the final RGB image $\mathbf{C}$.

Through the decoupling of geometry and appearance attributes, each Gaussian primitive exclusively stores 10 floating-point values for its explicit geometric representation, eliminating the need for storing high-overhead spherical harmonic (SH) features. Additionally, we employ a hash table with $L$ levels, where each level contains a table of size $T$ with $F$ feature dimensions per entry. The memory consumption of the hash table is $L\times T\times F$.

\setlength{\tabcolsep}{2.5pt}
\begin{table*}[!t]
\centering
\captionsetup{position=bottom}  
\caption{\textbf{Quantitative comparisons on the NeRFSyn, DTU, and MipNeRF360-indoor dataset.} We evaluate the rendering quality using PSNR$\uparrow$, SSIM$\uparrow$, and LPIPS$\downarrow$ metrics, and assess model compactness by reporting the average number of Gaussian points and model size. We report our model size with both the storage of 2D Gaussians and hash table feature vectors.}
\resizebox{1.0\textwidth}{!}{
\begin{tabular}{l|ccccc|ccccc|ccccc}
\toprule
 & \multicolumn{5}{c|}{NeRFSyn} & \multicolumn{5}{c|}{DTU} & \multicolumn{5}{c}{MipNeRF360-indoor}\\ \hline
 & PSNR $\uparrow$  & SSIM $\uparrow$ & LPIPS $\downarrow$ & Points $\downarrow$ & Size $\downarrow$  & PSNR $\uparrow$  & SSIM $\uparrow$ & LPIPS $\downarrow$ & Points $\downarrow$ & Size $\downarrow$  & PSNR $\uparrow$  & SSIM $\uparrow$ & LPIPS $\downarrow$ & Points $\downarrow$ & Size $\downarrow$ \\ \hline

3DGS & 33.34 & \cellcolor{Snd}0.969 & \cellcolor{Fst}0.030 & 288k & 68MB & 33.77 & 0.965 & \cellcolor{Trd}0.044 & 359k & 85MB & \cellcolor{Fst}31.03 & \cellcolor{Fst}0.921 & \cellcolor{Snd}0.188 & 1457k & 344MB\\ 
2DGS & 33.15 & \cellcolor{Trd}0.968 & 0.034 & \cellcolor{Trd}102k & \cellcolor{Fst}24MB & 33.89 & \cellcolor{Snd}0.966 & 0.048 & \cellcolor{Trd}214k & \cellcolor{Snd}47MB & 30.29 & \cellcolor{Snd}0.920 & 0.189 & \cellcolor{Trd}876k & \cellcolor{Snd}204MB\\ 
SuperGS & \cellcolor{Fst}33.71 & \cellcolor{Fst}0.970 & \cellcolor{Snd}0.031 & 207k & 69MB & \cellcolor{Trd}33.94 & \cellcolor{Fst}0.967 & \cellcolor{Snd} 0.043 & 394k & 132MB & 30.23 & \cellcolor{Trd}0.917 & \cellcolor{Snd}0.188 & 1316k & 463MB\\
GsTex& \cellcolor{Trd}33.37 & 0.965 & 0.041 & \cellcolor{Snd}100k & \cellcolor{Trd}38MB & \cellcolor{Fst}33.98 & 0.964&  0.045&  \cellcolor{Snd}186k& \cellcolor{Trd}61MB & \cellcolor{Trd}30.46 & 0.915 & 0.204 & \cellcolor{Snd}784k & \cellcolor{Trd}221MB\\
Ours & \cellcolor{Snd}33.50 & 0.967 & \cellcolor{Trd}0.032 & \cellcolor{Fst}73k & \cellcolor{Snd}2+28MB & \cellcolor{Snd}33.96 & \cellcolor{Trd}0.965 &  \cellcolor{Fst}0.042 & \cellcolor{Fst}80k & \cellcolor{Fst}3+28MB & \cellcolor{Snd}30.59 & 0.911 & \cellcolor{Fst}0.174 & \cellcolor{Fst}356k & \cellcolor{Fst}13+168MB \\
\bottomrule
\end{tabular}
}
\label{tab:nvs-comparison}
\end{table*}

\subsection{Optimization}
To accelerate training and ensure more stable convergence, we draw inspiration from~\cite{li2023neuralangelo} and employ a feature level annealing strategy:
\begin{equation}
    \mathbf{f}(\mathbf{x},\lambda) = (w_i(\lambda)\mathbf{f}^{i}(\mathbf{x})),  \hspace{0.2cm} w_i(\lambda) = 
\begin{cases} 
0 & \text{if } i > \lambda \\
1 & \text{otherwise}
\end{cases}
\end{equation}
The parameter $\lambda$  controls the number of active hash encoding levels, $i$ is the level index of hash grid. 



Instead of computing 3D world-space points using the depth of intersections and ray direction, we use the splat-to-world transformation matrix: $\mathbf{p}_\mathbf{u}=\mathbf{H}(u,v,1,1)^{\mathrm{T}}$, which leads to the following gradients with respect to the intersection $\mathbf{u}$ and the homogeneous matrix $\mathbf{H}$: 
\begin{equation}
\begin{split}
    \frac{\partial L}{\partial \mathbf{u}} = \frac{\partial L}{\partial \mathcal{G}}\frac{\partial \mathcal{G}}{\partial \mathbf{u}} + &\frac{\partial L}{\partial d}\frac{\partial d}{\partial \mathbf{u}}+
    \frac{\partial L}{\partial \mathbf{f}}\frac{\partial \mathbf{f}}{\partial \mathbf{p}_\mathbf{u}} 
    \frac{\partial \mathbf{p}_\mathbf{u}}
    {\partial \mathbf{u}},  \hspace{0.4cm}
    \\
    \frac{\partial L}{\partial \mathbf{H}} = &\frac{\partial L}{\partial \mathbf{f}}\frac{\partial \mathbf{f}}{\partial \mathbf{p}_\mathbf{u}} \frac{\partial \mathbf{p}_\mathbf{u}}{\partial \mathbf{H}}. 
\end{split}
\end{equation}
where $d$ is the depth of ray-splat intersection and the first two terms in $\frac{\partial L}{\partial \mathbf{u}}$ originate from 2DGS.

We optimize our model to minimize the following loss:
\begin{equation}
    \mathcal{L} = \mathcal{L}_c + \alpha \mathcal{L}_d + \beta \mathcal{L}_n + \gamma\mathcal{L}_a.
\end{equation}
where $\mathcal{L}_c$ is an RGB loss combining $\mathcal{L}_1$ and $\mathcal{L}_{ssim}$ used in 3DGS~\cite{kerbl3Dgaussians}, $\mathcal{L}_d$ and $\mathcal{L}_n$ are regularization terms used in 2DGS~\cite{Huang2DGS2024}, and $\mathcal{L}_a$ is an alpha map loss.
Following 2DGS, we set $\alpha = 1000$ for bounded scenes, $\alpha = 100$ for unbounded scenes, and $\beta = 0.05$ for all scenes. 
The coarse-to-fine parameter $\lambda$ is incremented every 3,000 iterations.

\begin{figure*}[t] 
\centering
\includegraphics[width=1\linewidth]{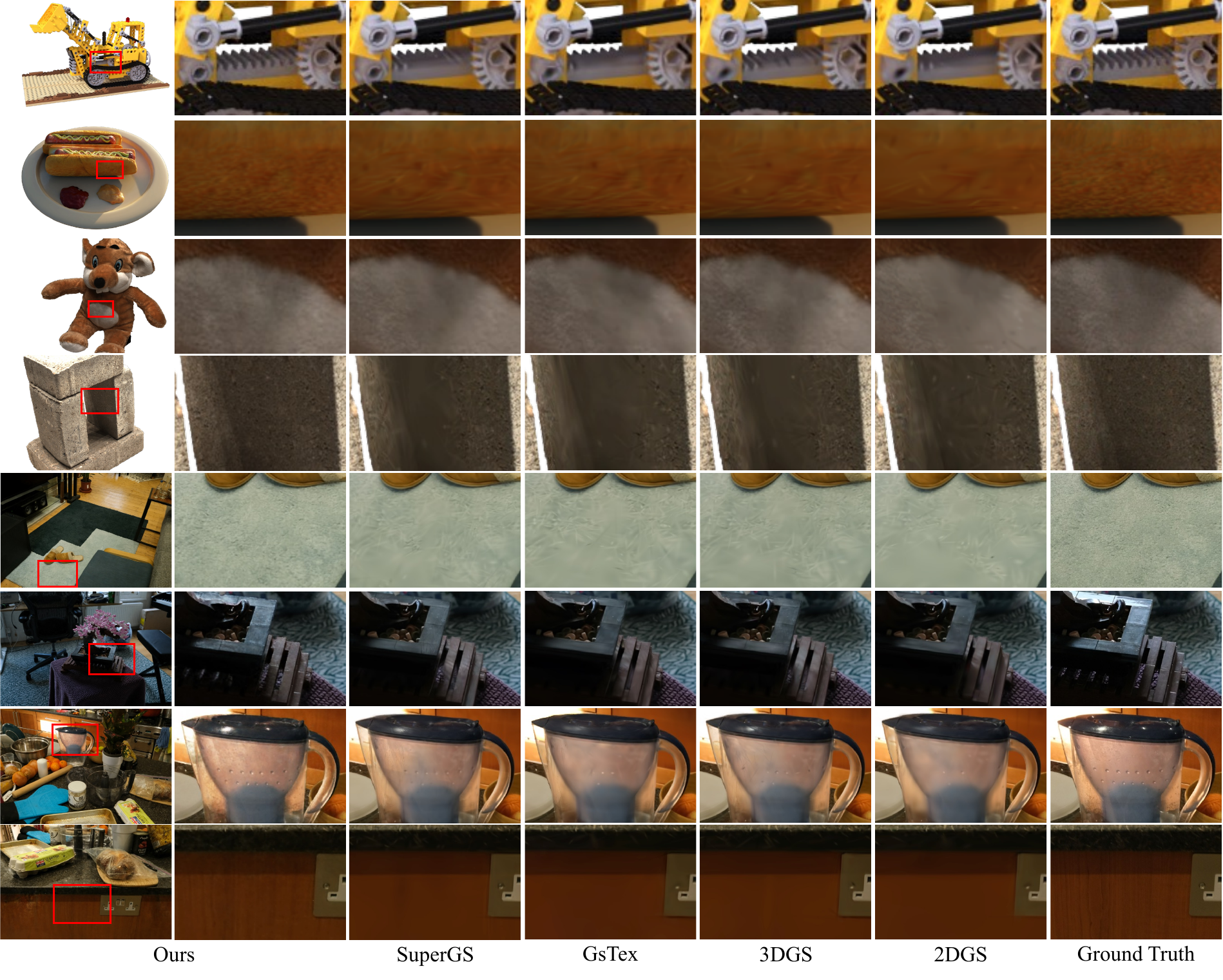}
\caption{\textbf{Qualitative comparisons on the NeRFSyn, DTU, and MipNeRF360-indoor dataset.} Our method consistently recovers clearer details on texture-rich regions across different scenes. \it{Zoom-in for best visualization}.}
\label{fig:cmp_main} 
\end{figure*}
\begin{figure*}[t] 
\centering
\includegraphics[width=1\linewidth]{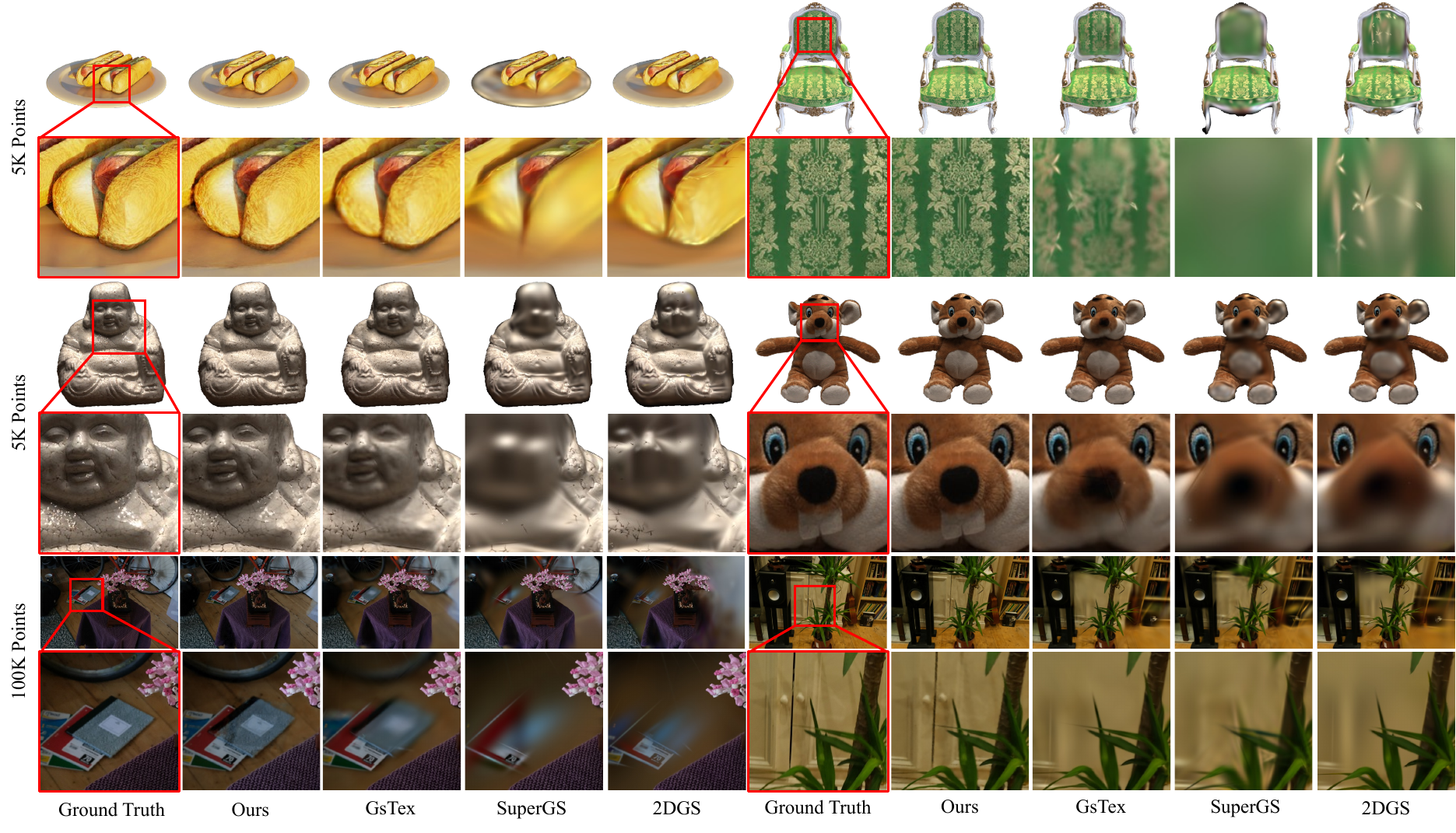}
\caption{\textbf{Qualitative comparisons on reconstruction using fewer Gaussian points.} Each row shows the results with the number of points limited to 5k for object, and 100k for scene, respectively. Our method significantly outperforms existing baselines due to the decoupling of geometry and appearance.}
\label{fig:cmp_fewer} 
\end{figure*}
\begin{figure*}[t] 
\centering
\includegraphics[width=1\linewidth]{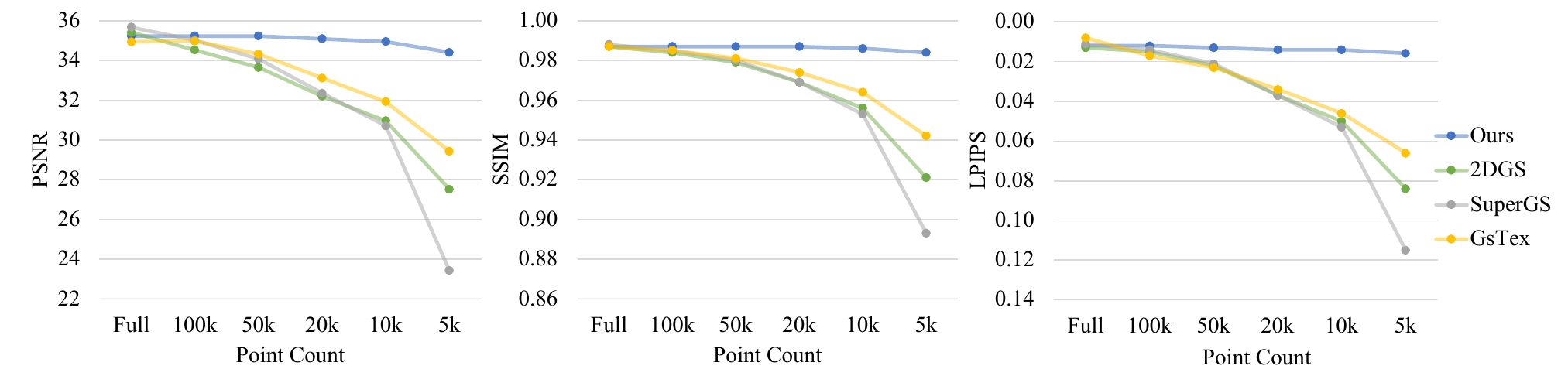}
\caption{\textbf{Quantitative comparisons on reconstruction using fewer Gaussian points.} We report the PSNR, SSIM, and LPIPS on the NeRFSyn Chair scene. Our method maintains high-quality rendering with a significantly smaller number of points compared to other methods.}
\label{fig:less_points} 
\end{figure*}

\begin{figure}[t] 
\centering
\includegraphics[width=0.8\linewidth]{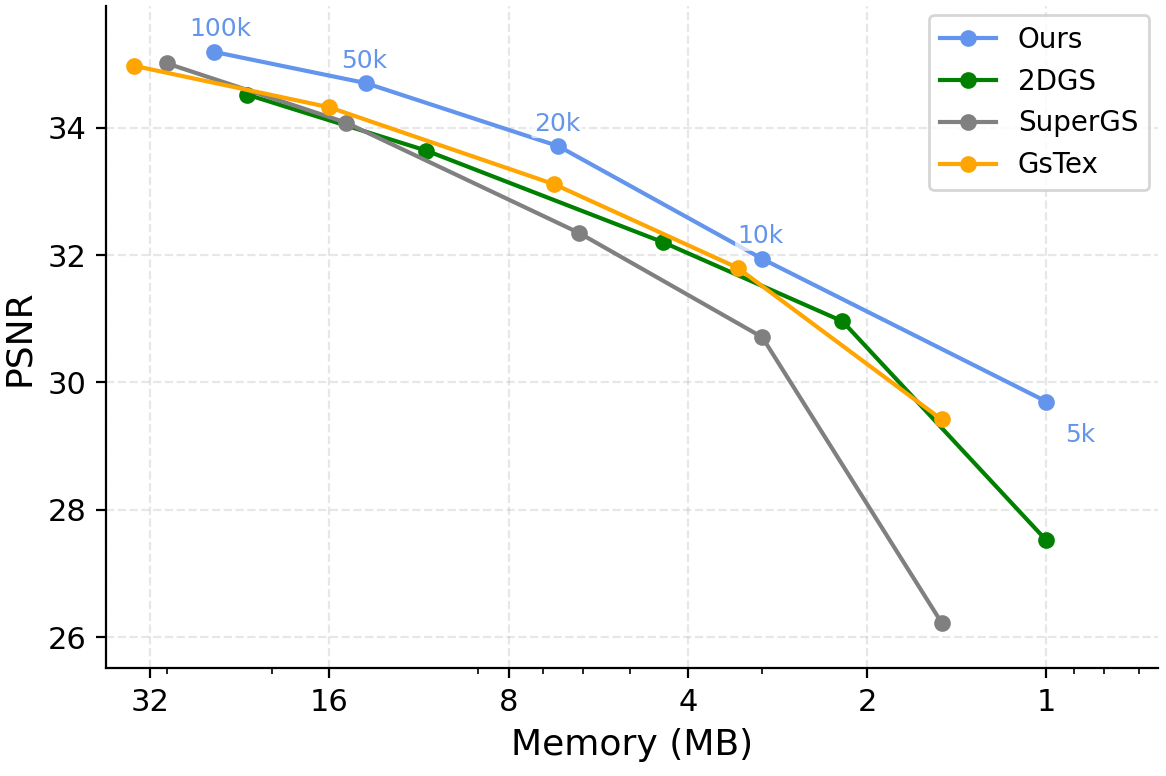}
\caption{\textbf{Quantitative comparisons with respect to model size using fewer Gaussian primitives.} The number of primitives is set to 100k, 50k, 20k, 10k, and 5k for all methods. Our method achieves higher texture quality while maintaining a compact overall model size.}
\label{fig:cmp_modelsize} 
\end{figure}

\section{Experiments}

\subsection{Implementation}

We implement our method based on the 2DGS framework. We extend the hash encoding in the CUDA kernel during rasterization and use an MLP to decode the final RGB map in PyTorch. 
To accelerate training, we first pre-train 2DGS for 10,000 iterations to obtain 2D splat initialization, followed by joint optimization of 2D splats, hash features, and the MLP for 20,000 iterations. All parameters are trained jointly and purely based on the rendering loss and regularization terms, without any alternating optimization.

For object-level datasets, we render the alpha maps of all training views at the end of 2DGS initialization as a constraint, with $\gamma = 0.1$ to regularize our alpha map.


We apply a gradient threshold of 4e-4 and reset opacity every 3,000 steps, disabling point pruning for 1,000 steps after each opacity reset to ensure training stability. For hash encoding hyper-parameters, we set level $L$ = 6, table size $T = 2^{19}$ to $2^{21}$, number of feature dimensions $F$ = 4. We also use contract function in MipNeRF360~\cite{barron2022mipnerf360} for scene dataset. All experiments are conducted on a single NVIDIA RTX 4090 GPU. For additional details, please refer to the Appendix.

\subsection{Comparison}

\noindent \textbf{Dataset.\label{Dataset}} 
We evaluate our method on the NeRFSynthetic dataset~\cite{mildenhall2020nerf}, DTU dataset~\cite{Aanes2016IJCV}, and Mip-NeRF 360 dataset~\cite{Barron2021ICCV} following established protocols \cite{Huang2DGS2024, wang2021neus}. 
We use PSNR, SSIM, and LPIPS to evaluate the quality of novel-view-synthesis (NVS) for all methods. For surface reconstruction, we measure geometric accuracy using Chamfer Distance (CD) on the DTU dataset.

\noindent \textbf{Novel  View Synthesis \label{NVS-compare}}
We compare our method with baselines including 2DGS~\cite{Huang2DGS2024}, 3DGS~\cite{kerbl3Dgaussians}, and the state-of-the-art methods SuperGS~\cite{xu2024supergaussians} and GsTex~\cite{rong2024gstex}, allowing spatially varying colors in Gaussian primitives. \tabref{tab:nvs-comparison} presents the PSNR, SSIM, LPIPS, number of Gaussian primitives, and model size across different datasets. Experimental results demonstrate that our method achieves comparable rendering quality  to state-of-the-art Gaussian-based approaches across multiple datasets, while requiring much fewer Gaussian points and maintaining a more compact model size. Notably, owing to enhanced representation of texture details, our method demonstrates superior performance on the LPIPS metric, which is highly correlated with human visual perception.

\cref{fig:cmp_main} presents qualitative comparisons of novel view synthesis across multiple datasets. In regions with high-frequency details, previous methods exhibit \emph{needle-like} artifacts due to excessive densification, resulting in numerous tiny Gaussian primitives.
Although SuperGS and GsTex enable spatially varying colors, they still fail to accurately reconstruct these fine details. In contrast, our method faithfully reproduces photorealistic texture details with enhanced visual fidelity.

\setlength{\tabcolsep}{7pt}
\begin{table}[htbp]
\centering
\caption{ \textbf{Scale Anisotropy Analysis.} We define the anisotropic scale ratio as $ \frac{\mathrm{max}(s_0,s_1)}{\mathrm{min}(s_0,s_1)}$, and call Gaussian with ratio $< 0.1$ as \emph{needle-like} Gaussian. We report these two ratios on the NeRFSyn dataset and show that our method produces more geometrically balanced primitives without shape regularization.}
\captionsetup{position=bottom}  
\label{tab:Needle table}
\resizebox{0.48\textwidth}{!}{
\begin{tabular}{l|cccc}
& Ours & SuperGS & GsTex & 2DGS \\ \hline
anisotropic scale ratio & 0.331 & 0.289 & 0.291 & 0.295 \\
\emph{needle-like} ratio & 0.156 & 0.209 & 0.230 & 0.206 \\
\end{tabular}
}
\end{table}


To further validate the impact of decoupling on texture enhancement, we compare the rendering quality of different methods under varying numbers of primitives. As shown in \cref{fig:cmp_fewer} and \cref{fig:less_points}, our method maintains consistent rendering quality with superior texture details across different primitive densities, demonstrating the robustness of our complete geometry-appearance decoupling approach. 

Unlike previous Gaussian-based methods, where model size scales linearly with the number of primitives, only the geometry footprint $(\textbf{p}, \textbf{S}, \textbf{R}, \alpha)$ in our decoupled model grows with the primitive count. Therefore, we show the trend of rendering quality with respect to model size under fewer Gaussian primitives in \cref{fig:cmp_modelsize}, achieved by adjusting our hash table size (directly reducing $L$ and $T$) to match the appearance overhead of 2DGS (\textbf{SH}).





We further conduct quantitative analysis by measuring scale anisotropy ratios and \emph{needle-like} primitive proportions at~\tabref{tab:Needle table}. Though SuperGS and GsTex utilize local color variations for improved appearance modeling, they exhibit similar anisotropy ratios to the 2DGS baseline.
Our method achieves more geometrically-balanced primitives and maintaining high-fidelity detail with fewer \emph{needle-like} artifacts, due to the full decoupling of geometry and appearance.

\begin{figure}[t] 
\centering
\includegraphics[width=1\linewidth]{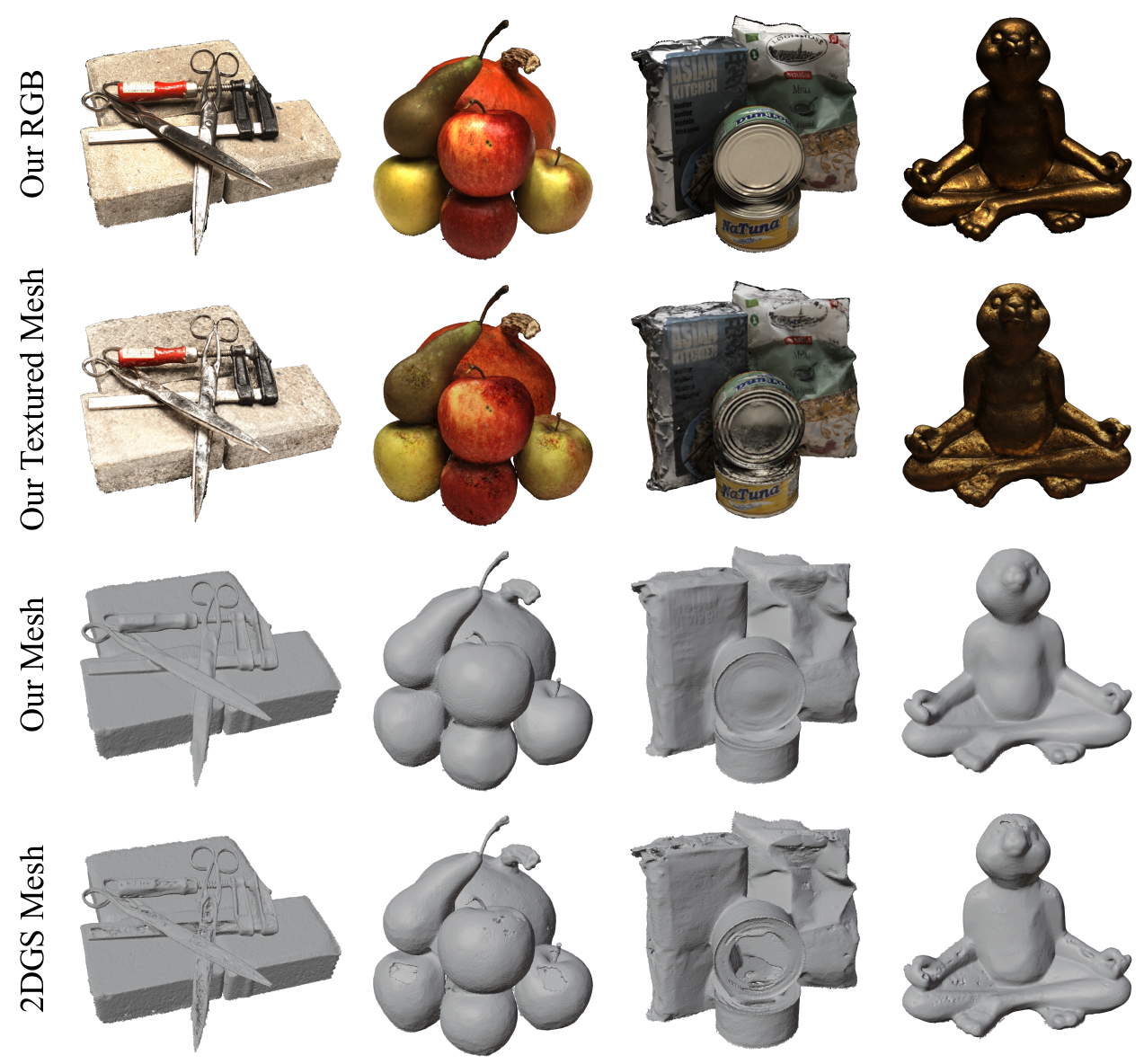}
\caption{\textbf{Qualitative comparison of geometry reconstruction on the DTU dataset.} Our decoupled representation makes the geometry robust to highly challenging view-dependent effects, producing noticeable improvements in reconstruction quality.}
\label{fig:cmp_mesh} 
\end{figure}

\noindent \textbf{Geometry Reconstruction. \label{geometry-compare}}
%
In~\tabref{tab:geometry}, we evaluate the quality of geometry reconstruction against the 2DGS~\cite{Huang2DGS2024}, GaussianSurfel~\cite{Dai2024GaussianSurfels}, 3DGS~\cite{kerbl3Dgaussians} and NeuS~\cite{wang2021neus} baselines using Chamfer distance on the DTU dataset. Our method achieves comparable reconstruction quality to 2DGS with significantly fewer points. We further visualize the rendering and mesh results in~\cref{fig:cmp_mesh}, demonstrating that our decoupling approach enables more stable reconstruction in highly view-dependent regions.


\setlength{\tabcolsep}{10pt}
\begin{table}[t!]
\centering
\caption{\textbf{Quantitative comparison of geometry reconstruction on the DTU dataset}. We report the Chamfer Distance (CD) in millimeters compared with baselines.}\label{tab:cmp_dtu}
\resizebox{0.48\textwidth}{!}{
\begin{tabular}{c|ccccc}
\toprule
Methods & Ours & 2DGS & GaussianSurfel & 3DGS & NeuS \\ \hline
CD $\downarrow$ & \cellcolor{Snd}0.83 & \cellcolor{Fst}0.80 & 0.88 & 1.96 & \cellcolor{Trd}0.84 \\
Points $\downarrow$ & \cellcolor{Fst}80k & \cellcolor{Trd}214k & \cellcolor{Snd}168k & 359k  & - \\
\bottomrule
\end{tabular}}
\label{tab:geometry}
\end{table}

\begin{figure}[t] 
\centering
\includegraphics[width=1\linewidth]{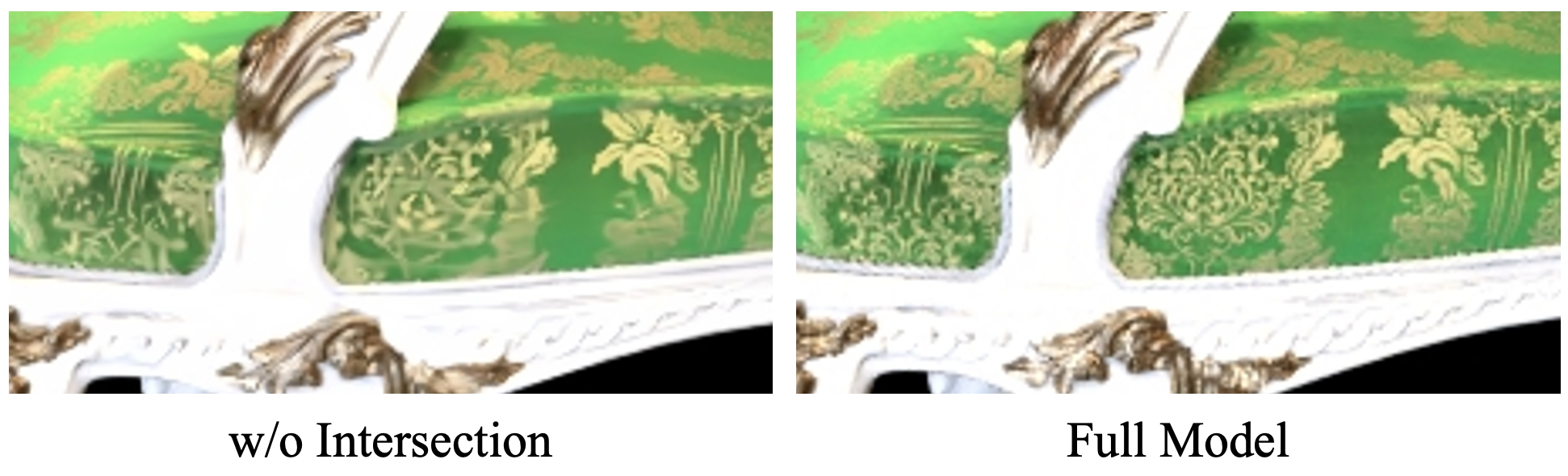}
\caption{\textbf{Qualitative results on spatially varying features.} Our method renders high-fidelity texture details and effectively mitigates the \emph{needle-like} artifacts.}
\label{fig:wo_inter} 
\end{figure}




\subsection{Ablation Study}

In this section, we conduct ablation studies to evaluate the impact of design choices on NVS quality using the NeRF Synthetic dataset and geometry reconstruction quality on the DTU dataset, with comprehensive quantitative results provided in~\tabref{tab:combined}.

\begin{table}[htbp]
\centering
\setlength{\tabcolsep}{1pt}
\captionsetup{position=bottom}
\caption{\textbf{Ablation study on model design.}} 
\label{tab:combined} 

\begin{subtable}[t]{0.26\textwidth} 
\centering
\resizebox{\textwidth}{!}{
\begin{tabular}{l|ccc}
\hline
 & PSNR$\uparrow$ & SSIM$\uparrow$ & LPIPS$\downarrow$ \\ \hline
w/o intersection  & 33.41 & \textbf{0.969} & 0.033  \\ 
Full Model & \textbf{33.50} & 0.967 & \textbf{0.032} \\
\hline
\end{tabular}
}
\subcaption{NVS quantities on NeRFSyn.} 
\label{tab:sub_view}
\end{subtable}
\hfill
\begin{subtable}[t]{0.21\textwidth}
\centering
\resizebox{\textwidth}{!}{
\begin{tabular}{l|cc}
\hline
 & CD$\downarrow$ & PSNR$\uparrow$ \\ \hline
A.w/ depth-ray & 0.89 & 33.82 \\ 
B.w/o $\partial\mathbf{F}/\partial\mathbf{x}$ & 0.95 & 33.60 \\ \hline
Full Model & \textbf{0.83} & \textbf{33.96} \\
\hline
\end{tabular}
}
\subcaption{Reconstruction qualities on DTU.} 
\label{tab:sub_geom}
\end{subtable}
\end{table}

\noindent \textbf{Spatially Varying Features}
We first validate the effectiveness of spatially varying features at each intersection $(u,v)$. As evidenced by ~\tabref{tab:sub_view} and ~\cref{fig:wo_inter}, replacing our $(u,v)$-based hash encoding $\mathbf{f}(\mathbf{H}(u_i,v_i,1,1)^{\mathrm{T}})$ with Gaussian center encoding $\mathbf{f}(\mathbf{p_i})$ significantly degrades the ability to represent high-frequency details.

\noindent \textbf{Coordinate Gradient Analysis}
We replace the homogeneous transformation-based mapping $\mathbf{x_i}=\mathbf{H}(u_i,v_i,1,1)^{\mathrm{T}}$ with depth-ray formulation $\mathbf{x}_i = z_i \cdot \mathbf{d} + \mathbf{o}$, which redirects gradient flow through depth. As shown in~\tabref{tab:sub_geom} A, this implementation restricts intersection position optimization, leading to degraded geometry quality.

We further disable the $\partial{\mathbf{F}}/\partial{\mathbf{x}}$ gradient term, which is absent in the original Instant-NGP framework, as no gradient propagates from features to ray sample positions. Experiments in~\tabref{tab:sub_geom} B demonstrate this gradient term effectively enhances Gaussian parameter optimization, directly improving reconstruction quality.

\setlength{\tabcolsep}{15pt}
\begin{table}[htbp]
\centering
\caption{\textbf{Training Time and FPS Comparison.} We compare training time (seconds) and rendering speed (FPS) on NeRFSyn dataset.}
\captionsetup{position=bottom}  
\label{tab:Time}
\resizebox{0.35\textwidth}{!}{
\begin{tabular}{l|cc}
& Training Time(s)$\downarrow$  & FPS$\uparrow$\\ \hline
2DGS  & 312 & 224 \\
Ours  & 1573 & 71 \\
\end{tabular}
}
\end{table}




\subsection{Explicit texture map extraction}

Thanks to the decoupling of geometry and texture, as well as our global feature field representation, we can bake the trained feature field into an explicit texture map. This operation treats the mesh surface as sampling plane with 1.0 opacity, so each position $\mathbf{x}$ corresponds with an explicit RGB color by $M(\mathbf{f}(\mathbf{x}), \mathbf{d})$. Specifically, after extracting the mesh corresponding to 2D primitives using the Mesh Extraction method in 2DGS, we unwrap the mesh to obtain an UV map. By using the world coordinates corresponding to each pixel in the UV map as sampling points, an explicit texture map can be generated through hash encoding and MLP. We present our textured-mesh results for NeRFSyn and DTU dataset in the second row of Fig. \ref{fig:cmp_mesh}.

\section{Conclusion}

We have introduced a simple yet fundamental adjustment to Gaussian Splatting by proposing a geometry-appearance disentanglement representation, termed NeST-Splatting. Our approach leverages a multi-level hash grid to model the texture field, while Gaussian primitives serve as samplers to splat the texture field into images.
Our representation enables high-fidelity texture reconstruction with significantly fewer primitives.
Extensive experiments on NVS quality and geometry reconstruction demonstrate the effectiveness and efficiency of our method.  
We hope our findings on disentangled representations will inspire further research in related tasks.

\noindent \textbf{Limitations.} 
Our method exhibits slower rendering and training speeds compared to Gaussian Splatting baseline due to the additional overhead of multi-resolution feature querying and MLP decoding (\tabref{tab:Time}). We aim to address this limitation in future work by exploring more efficient integration of explicit geometry representations.

\section*{Acknowledgments}



We thank the anonymous reviewers for their constructive comments. We also thank Xiuchao Wu, Hongyu Tao, Haoming Yu, and Jiamin Xu for their insightful discussions. Weiwei Xu is partially supported by NSFC grant No.~62421003 and National Key Research and Development Program of China No.~2024YFE0216600. This paper is supported Yongjiang Innovation Project No.~2025Z062, and the Information Technology Center and State Key Lab of CAD\&CG, Zhejiang University.

{
    \small
    \bibliographystyle{ieeenat_fullname}
    \bibliography{bibliography,bibliography_long,bibliography_custom}
}

\maketitlesupplementary
\appendix


\section{Implementation Details}

\noindent \textbf{Hash Encoding Parameters.}
We conducted experiments on the hash encoding parameters $F,L,T$ using the NeRFSyn dataset in~\tabref{tab:hash_param}:

\begin{table}[htbp]
\setlength{\tabcolsep}{12pt} 
\small 
\centering
\caption{Ablation study on hash grid parameters $F,L,T$.}
\captionsetup{position=bottom, skip=-3pt, font=tiny} %
\label{tab:hash_param}
\resizebox{0.45\textwidth}{!}{ 
\begin{tabular}{c|ccc}
$F=2,T=2^{19}$ & $L=8$& $L=12$ & $L=16$\\
PSNR$\uparrow$ & \textbf{33.09} & 33.09 & 32.94 \\ \hline
$F=4,T=2^{19}$ & $L=4$ & $L=6$ & $L=8$ \\
PSNR$\uparrow$ & 33.30 & \textbf{33.50} & 33.28\\ \hline
$F=4,L=6$ & $T=2^{18}$ & $T=2^{19}$ & $T=2^{20}$ \\
PSNR$\uparrow$ & 33.37 & \textbf{33.50} & 33.50 \\
\end{tabular}
}
\end{table}
Considering the trade-offs between memory efficiency and reconstruction quality, we selected $F=4$ and $L=6$ as the main settings. For $T$, we used $2^{19}$ for object-level data and $2^{21}$ for scene-level data in our experiments. 

\noindent \textbf{Dataset Setting.}
The NeRFSyn dataset consists of 8 synthetic scenes at a resolution of $800\times800$. The DTU dataset includes 15 scenes, each with 49 or 64 images at a resolution of $1600\times1200$. Following 2DGS, we use Colmap sparse points and train at a reduced resolution of $800\times600$ for efficiency. We evaluate the DTU dataset using a fixed and consistent evaluation protocol, selecting images with indices $8, 13, 16, 21, 26, 31$, and $34$. If the number of images exceeds $56$, we additionally include the image with index $56$.
The MipNeRF360 dataset includes 5 outdoor and 4 indoor scenes. We train and test at half resolution for indoor scenes and quarter resolution for outdoor scenes, with test views sampled every 8 images. 

\noindent \textbf{Training Details}
The training process for GsTex involves two stages: we first train 2DGS from scratch for 15,000 iterations to obtain an initial set of Gaussians, followed by training GsTex for an additional 15,000 iterations. For SuperGS, we did not impose a restriction on the growth in the number of Gaussians, as we found that an inappropriate upper limit could lead to suboptimal results. All of our training was conducted on a single NVIDIA RTX 4090 GPU with 24GB of memory, which can accommodate a maximum of approximately 5.7 million Gaussians. Consequently, we encountered out-of-memory errors on the bicycle and treehill scenes from the MipNeRF360 dataset.

\section{MipNeRF360 Results}

\noindent \textbf{Contraction Function}
As proposed in Mip-NeRF360, We map unbounded background into a bounded cubic region using the following contraction function:
\begin{equation}
    \mathrm{contract}(\mathbf{x}) = 
\begin{cases} 
 \mathbf{x} & \left\|\mathbf{x}\right\| \leq 1 \\
(2-\frac{1}{\left\|\mathbf{x}\right\|})(\frac{\mathbf{x}}{\left\|\mathbf{x}\right\|}) & \left\|\mathbf{x}\right\|> 1
\end{cases}
\end{equation}
The coordinates of any ray-splat intersection is first normalized and then contracted before querying the hash grid features. The entire scene is contracted into a bounded $[-2, 2]$, with the foreground region normalized to the $[-1, 1]$. 

\setlength{\tabcolsep}{0.5pt}
\begin{table}[htbp]
\centering
\captionsetup{position=bottom}  
\resizebox{0.475\textwidth}{!}{
\begin{tabular}{l|ccccc|ccccc}
& \multicolumn{5}{c|}{Outdoor Scene} & \multicolumn{5}{c}{Indoor Scene} \\
& PSNR$\uparrow$ & SSIM$\uparrow$ & LPIPS$\downarrow$ & Points$\downarrow$ & Size$\downarrow$  & PSNR$\uparrow$ & SSIM$\uparrow$ & LPIPS$\downarrow$ & Points$\downarrow$  & Size$\downarrow$\\ \hline
3DGS & \cellcolor{Snd}24.24 & \cellcolor{Trd}0.704 & \cellcolor{Trd}0.283 & 4821k & 1140MB & \cellcolor{Fst}31.03 & \cellcolor{Fst}0.921 & \cellcolor{Snd}0.188 & 1457k & 344MB \\
2DGS & \cellcolor{Fst}24.33 & \cellcolor{Fst}0.708 & \cellcolor{Trd}0.283 & \cellcolor{Snd}3360k & \cellcolor{Trd}782MB & 30.29 & \cellcolor{Snd}0.920 & 0.189 & \cellcolor{Trd}876k & \cellcolor{Snd}204MB\\
SuperGS & OOM & OOM & OOM & OOM & OOM & 30.23 & \cellcolor{Trd}0.917 & \cellcolor{Snd}0.188 & 1316k & 463MB \\
GsTex & \cellcolor{Snd}24.24 & \cellcolor{Fst}0.708 & \cellcolor{Snd}0.276 & \cellcolor{Trd}3067k & \cellcolor{Snd}663MB & \cellcolor{Trd}30.46 & 0.915 & 0.204 & \cellcolor{Snd}784k & \cellcolor{Trd}221MB\\
Ours & 23.85 & 0.690 & \cellcolor{Fst}0.257 & \cellcolor{Fst}1650k & \cellcolor{Fst}257MB & \cellcolor{Snd}30.59 & 0.911 & \cellcolor{Fst}0.174 & \cellcolor{Fst}356k & \cellcolor{Fst}181MB\\
\end{tabular}
}
\caption{\textbf{Quantitative results on MipNeRF360 indoor and outdoor scenes.} OOM indicates out-of-memory on a 24GB GPU. Our method achieves significant improvements in LPIPS scores while maintaining a smaller number of Gaussian primitives and a more compact model size.}
\label{tab:mip-all}
\end{table}

\begin{figure*}[t!]
\begin{center}
\includegraphics[width=0.95\linewidth]{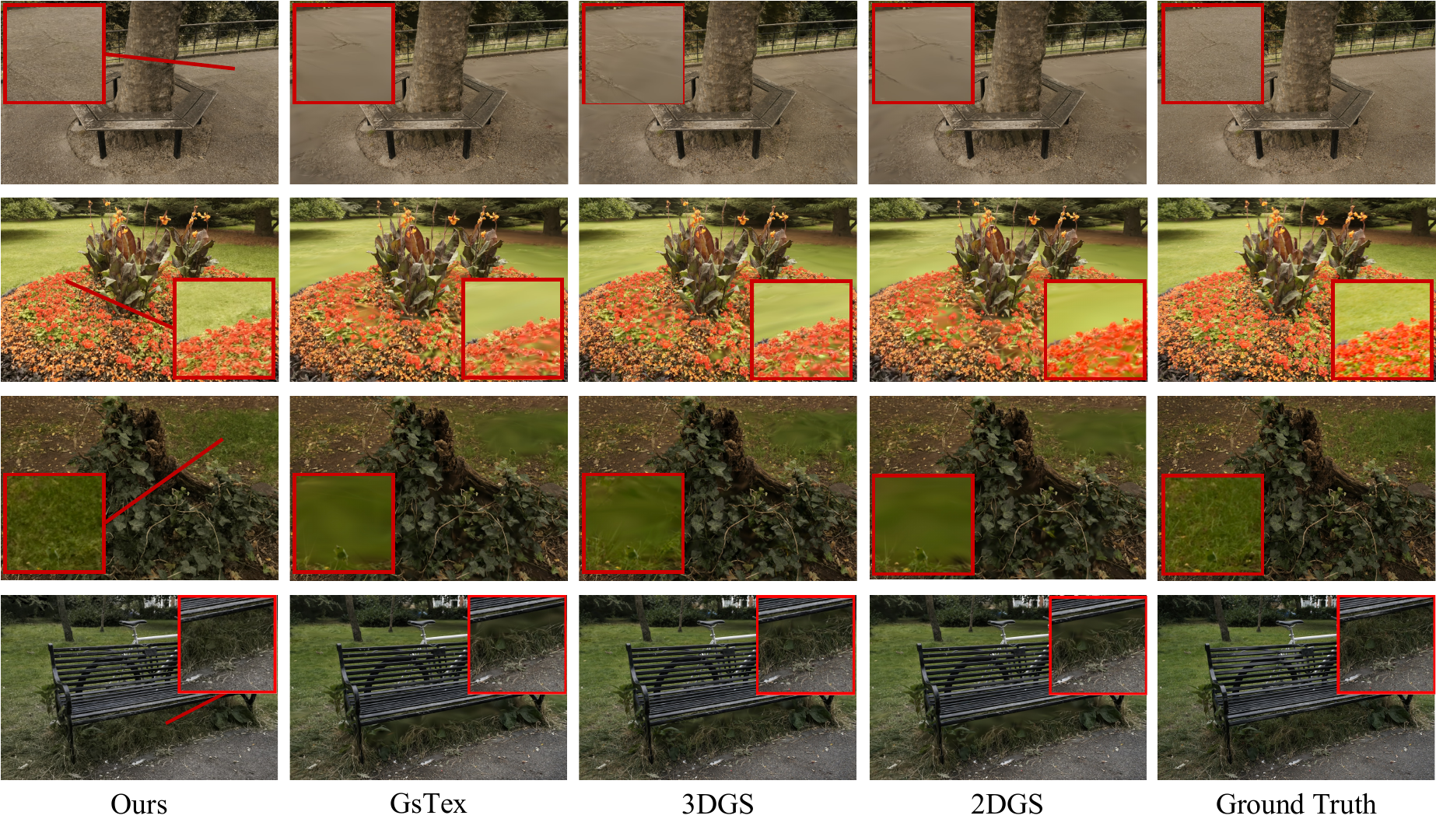}
\captionof{figure}{\textbf{Qualitative comparisons on the MipNeRF360-outdoor dataset.} Our method reveals finer details without densifying a large number of Gaussian primitives.}
\label{fig:360_out_cmp}
\end{center}
\end{figure*}

\noindent \textbf{Outdoor Scene}
We report all metrics for both indoor and outdoor scenes in~\tabref{tab:mip-all}. Our method produces more detailed rendering results and significantly improves the LPIPS metric, which aligns well with human visual perception. However, our method tends to overfit under-constrained background regions in outdoor scenes, resulting in lower PSNR and SSIM scores. We show qualitative comparisons in~\figref{fig:360_out_cmp}, where our method achieves photo-realistic rendering results, particularly on flat, texture-rich regions such as the ground and grass without requiring the densification of a large number of Gaussian primitives.

\section{More Results}


We present detailed quantitative results of all methods on the NeRFSyn, DTU, and MipNeRF360 datasets in~\tabref{tab:nerfsyn_cmp},~\tabref{tab:dtu_cmp}, and~\tabref{tab:360_cmp}, reporting PSNR, SSIM, and LPIPS metrics. We also invite readers to refer to our video results for better visualization.

\vspace{0.4cm}




\setlength{\tabcolsep}{7pt}
\begin{table*}[t]
\centering
\captionsetup{position=bottom}  
\resizebox{0.8\textwidth}{!}{
\small
\begin{tabular}{l cccccccc c}
 \hline
 Method & Mic & Chair & Ship & Materials & Lego & Drums & Ficus & Hotdog & Mean \\ \hline
3DGS & 35.42 & 35.90 & 30.90 & 30.00 & 35.78 & 26.16 & 34.85 & 37.70 & 33.34 \\ 
2DGS & 35.20 & 35.41 & 30.66 & 29.74 & 35.25 & 26.12 & 35.39 & 37.47 & 33.15  \\ \cline{2-10}
SuperGS & 36.09 & 35.67 & 31.68 & 30.35 & 35.65& 26.32 & 36.10 & 37.82 & 33.71\\
GsTex & 34.78 & 35.23 & 30.78 & 30.29 & 35.82 & 26.12 & 35.96 & 37.93 & 33.37\\
Ours &  36.30 & 35.23 & 31.27 & 29.70 & 35.27 & 26.16 & 36.23 & 37.80 & 33.50 \\ \hline
3DGS & 0.992 & 0.987 & 0.907 & 0.960 & 0.983 & 0.955 & 0.987 & 0.985 & 0.969 \\ 
2DGS & 0.991 & 0.987 & 0.903 & 0.958 & 0.981 & 0.954 & 0.988 & 0.985 & 0.968 \\ \cline{2-10}
SuperGS & 0.992 & 0.988 & 0.909 & 0.959 & 0.981 & 0.955 & 0.988 & 0.985 & 0.970\\
GsTex & 0.975 & 0.986 & 0.892 & 0.958 & 0.981 & 0.952 & 0.987 & 0.985 & 0.965\\
Ours & 0.992 & 0.987 & 0.898 & 0.955 & 0.980 & 0.950 & 0.988 & 0.987 & 0.967 \\ \hline
3DGS & 0.006 & 0.010 & 0.106 & 0.036 &  0.016 & 0.036 & 0.011 & 0.019 & 0.030 \\ 
2DGS & 0.007 & 0.013 & 0.117 & 0.040 &0.020 & 0.039 & 0.012 &
0.023 & 0.034 \\ \cline{2-10}
SuperGS & 0.007 & 0.011 & 0.103 & 0.037 & 0.018 & 0.037 & 0.011 & 0.020 & 0.031 \\
GsTex & 0.018 & 0.015 & 0.138 & 0.048 & 0.022 & 0.047 & 0.014 & 0.027 & 0.041\\
Ours & 0.007 & 0.012 & 0.097 & 0.044 & 0.018 & 0.047 & 0.012 & 0.019 & 0.032 \\ \hline
\end{tabular}
}
\caption{\textbf{Quantitative comparison of different methods on the NeRFSyn dataset.} We report PSNR$\uparrow$, SSIM$\uparrow$, and LPIPS$\downarrow$ scores.}
\label{tab:nerfsyn_cmp}
\end{table*}


\setlength{\tabcolsep}{3pt}
\begin{table*}[t]
\centering
\captionsetup{position=bottom}  
\resizebox{\textwidth}{!}{
\small
\begin{tabular}{l ccc cccc cccc cccc c}
 \hline
 Method & 24 & 35 & 40 & 55 & 63 & 65 & 69 & 83 & 97 & 105 & 106 & 110 & 114 & 118 & 122 & Mean \\ \hline
3DGS & 30.54 & 27.72 &30.91 &33.02 &36.47 &32.74 &30.43 &38.13 &29.52 &35.02 &36.27 &36.17 &31.87 &38.99 &38.69 &33.77 \\ 
2DGS & 30.68 & 27.78 & 31.26 & 33.28 & 35.67 & 33.21 & 30.59 & 37.85& 29.54& 34.78 & 36.48 & 35.86 & 32.32 & 39.57 & 39.50 & 33.89 \\ \cline{2-17}
SuperGS & 30.00 &27.92 & 31.20 & 33.63 & 35.69 & 32.72& 30.88& 37.38&29.94 & 34.39 & 36.77& 36.25& 32.55& 39.98& 39.73 & 33.94 \\
GsTex & 31.01 & 28.01 & 31.08 & 33.38 & 35.21 & 33.07 & 30.83 & 38.00 & 29.64 & 34.73 & 36.68 & 36.12 & 32.54 & 39.78 & 39.63 & 33.98\\
Ours & 30.37 & 27.84 &31.08 &33.26 &35.63 &32.94 &30.73 & 37.43 & 29.78 &34.82 & 36.57 & 36.52 & 32.45 & 39.95 & 40.04 &33.96 \\ \hline

3DGS & 0.946 & 0.933 &0.933 &0.975 &0.974 &0.972 &0.950 &0.983& 0.953 &0.969 &0.973 &0.976 &0.960 &0.981 &0.983 & 0.965 \\ 
2DGS & 0.946& 0.936 & 0.940 & 0.977 & 0.973& 0.975 &0.951 &0.982 &0.953 & 0.966 &0.974 & 0.974 & 0.963 & 0.981 & 0.985 & 0.966 \\ \cline{2-17}
SuperGS & 0.947& 0.938&0.942 &0.979 & 0.975 & 0.975 & 0.954 & 0.982 & 0.956 & 0.969 & 0.975 & 0.976 & 0.965 & 0.983 & 0.986 & 0.967\\
GsTex & 0.946 & 0.936 & 0.934 & 0.978 & 0.972 & 0.972 & 0.950 & 0.983 & 0.953 & 0.967 & 0.973 & 0.975 & 0.963 & 0.980 & 0.985 & 0.964\\
Ours & 0.944 & 0.940 & 0.937 & 0.976& 0.972 & 0.974 & 0.952&0.982 & 0.954 & 0.966 & 0.975 & 0.977 & 0.962 & 0.982 & 0.985 & 0.965  \\ \hline
3DGS & 0.045 &0.056 &0.089 &0.027 &0.029& 0.037 &0.062& 0.026 &0.056 &0.042 &0.044 &0.053 &0.044 &0.028 &0.020 &0.044 \\ 
2DGS & 0.053 & 0.055 & 0.086 & 0.025 & 0.031 & 0.039 & 0.066 & 0.030 & 0.062 & 0.053 & 0.046 & 0.060 & 0.051 & 0.033 & 0.022 & 0.048 \\ \cline{2-17}
SuperGS & 0.043 & 0.053 & 0.080 & 0.024 & 0.028 & 0.037 & 0.060 & 0.029 & 0.057 & 0.044 & 0.041 & 0.050 & 0.045 & 0.029 & 0.019 & 0.043\\
GsTex & 0.050 & 0.053 & 0.091 & 0.025 & 0.031 & 0.038 & 0.064 & 0.027 & 0.060 & 0.047 & 0.044 & 0.053 & 0.049 & 0.029 & 0.020 & 0.045\\
Ours & 0.046 & 0.054 & 0.081 & 0.027 & 0.029& 0.037& 0.059& 0.022& 0.056& 0.042 & 0.037 & 0.039 & 0.046 &  0.028 & 0.019 &  0.042\\ \hline
\end{tabular}
}
\caption{\textbf{Quantitative comparison of different methods on the DTU dataset.} We report PSNR$\uparrow$, SSIM$\uparrow$, and LPIPS$\downarrow$ scores.}
\label{tab:dtu_cmp}
\end{table*}

\vspace{2cm}
\setlength{\tabcolsep}{7pt}
\begin{table*}[t]
\centering
\captionsetup{position=bottom}  
\resizebox{0.9\textwidth}{!}{
\small
\begin{tabular}{l cccccc|ccccc}

 & \multicolumn{6}{c}{Outdoor Scene}& \multicolumn{5}{c}{Intdoor Scene}\\
 Method & bicycle & flowers & garden & stump & treehill & Mean & room & counter & kitchen & bonsai & Mean \\ \hline
3DGS &  24.71 &  21.09 &  26.63 & 26.45 & 22.33 & 24.24 & 31.50 & 28.96 & 31.38 & 32.26 & 31.03 \\ 
2DGS &  24.82 &  20.99 & 26.91 & 26.41 & 22.52 & 24.33 & 30.87 &  28.16 & 30.66 & 31.45 & 30.29\\ \cline{2-12}
SuperGS & OOM& 21.68& 27.31& 26.72& OOM& -& 30.00 & 28.71 & 30.66 & 31.57 & 30.23\\
GsTex &  24.68 & 21.17 & 26.76 & 26.24 & 22.33 & 24.24& 31.15 & 28.50 & 30.72 & 31.48 & 30.46\\
Ours & 24.49 & 20.05 & 26.68 & 25.87 & 22.20 & 23.85 & 31.30 & 28.45 & 30.61 & 32.02 & 30.59   \\ \hline
3DGS &  0.729 & 0.571 & 0.834 & 0.762 & 0.627 & 0.704 & 0.915 & 0.905 & 0.924 & 0.939 & 0.921 \\ 
2DGS &  0.731 & 0.573 & 0.845 & 0.764 & 0.630 & 0.708  & 0.915 & 0.905 &  0.924 &  0.939 & 0.920 \\ \cline{2-12}
SuperGS & OOM& 0.616& 0.867& 0.779& OOM& -& 0.908 & 0.905 & 0.922 & 0.931 & 0.917\\
GsTex & 0.730 & 0.582 & 0.849 & 0.758 & 0.621 &  0.708& 0.910 & 0.896 & 0.919 & 0.934 & 0.915 \\
Ours & 0.729 & 0.521 & 0.844 & 0.737 & 0.619 & 0.690 & 0.909 & 0.888 & 0.916 & 0.931 & 0.911 \\ \hline
3DGS & 0.265 & 0.377 & 0.147 & 0.266 & 0.362 & 0.283 & 0.219 & 0.201 & 0.127 & 0.205 & 0.188 \\ 
2DGS &  0.271 & 0.378 & 0.138 & 0.263 & 0.369 & 0.283  & 0.219 &  0.201& 0.127 &  0.205 & 0.189\\ \cline{2-12}
SuperGS & OOM& 0.320& 0.104& 0.020& OOM& -& 0.219 & 0.200 & 0.131 & 0.202 & 0.188  \\
GsTex & 0.265 & 0.365 & 0.138 & 0.248 & 0.365 &  0.276& 0.237 & 0.221 & 0.141 & 0.218 & 0.204\\
Ours & 0.236 & 0.358 & 0.129 & 0.246 & 0.317 &0.257 &0.194 & 0.202 & 0.125 & 0.176 & 0.174\\ \hline
\end{tabular}
}
\caption{ 
\textbf{Quantitative comparison of different methods on the MipNeRF360 dataset.} We report PSNR$\uparrow$, SSIM$\uparrow$, and LPIPS$\downarrow$ scores for both indoor and outdoor scenes.}
\label{tab:360_cmp}
\end{table*}

\end{document}